\documentclass[aps,prd,twocolumn,groupedaddress]{revtex4}
\usepackage{graphicx}
\usepackage{dcolumn}
\usepackage{bm}
\begin{document}

\title{A flux tube model for glueballs}

\author{Masaharu Iwasaki$^1$, Shin-Ichi Nawa$^2$, Takayoshi Sanada$^1$, and Fujio Takagi$^3$}
\affiliation{$^{1}$Department of Physics, Kochi University, Kochi 780-8520, Japan \\
$^{2}$Faculty of Human Life and Environmental Science, Kochi Women's University, Kochi 780-8515, Japan \\
$^{3}$Department of Basic Science, Ishinomaki Senshu University, Ishinomaki 986-8580, Japan}

\date{\today}

\begin{abstract}
We calculate the mass spectrum and the decay widths of glueballs in the flux tube model. The glueball is assumed to be a closed flux tube. The breathing motion and the rotational motion are investigated using the WKB approximation. The calculated spectra is consistent with those by lattice QCD. The decay widths are also computed using Schwinger mechanism and it is shown that they have rather large values.
\end{abstract}


\maketitle


\section{Introduction}

It is widely believed that the strong interaction is described by quantum chromodynamics (QCD) \cite{W80}. It is, however, much difficult to solve QCD so that several effective models are devised to explain physical properties of many hadrons. The flux tube model \cite{IP83}, or hadron string model \cite{MKB89}-\cite{M90}, is one of such models. According to the model, mesons are made up of a quark and an anti-quark connected by color flux tube and baryons are described by a quark and a diquark connected by the same flux tube. We showed in a previous paper \cite{IT99} that various excited states of the hadrons are explained systematically using such a simple model. On the other hand, the recent lattice QCD simulations \cite{MP99} predict the existence of glueballs, which are composed of only gluons . Unfortunately such a hadron has not been identified yet experimentally but the existence seems to be probable in Nature.

In this paper, we present an intuitive and analytic approach to the glueballs using a flux tube model. It is assumed that the glueball is a closed color flux tube (flux tube ring). Moreover the flux tube that constitutes the ring is assumed to have the same properties as the flux tube that constitutes a meson or a baryon. Consequently our model has no free parameter. We study the relativistic motion of the closed string and obtain the mass spectra of glueballs. Their decay widths are also calculated by taking into account $q\bar{q}$ pair production inside the flux tube (Schwinger mechanism).
This simple picture has been considered by several authors. In particular, Koma, Suganuma and Toki \cite{KST99} studied relativistic breathing motion analytically with the use of the dual Ginzburg-Landau theory \cite{S88}-\cite{SST95}. It is, however, very interesting to investigate the rotational motion of the closed string, which corresponds to the Regge trajectory of glueballs.

This paper is organized as follows. In the next section, the closed flux tube is formulated and we obtain a Lagrangian of the string. In Sec.3, canonical quantization is applied to the dynamical system and we obtain the Schr\"odinger equation. Then the mass spectra of the glueballs are obtained with the help of the WKB method. The decay widths of the glueballs are calculated by taking into account the quark pair creation mechnism in Sec.4. The numerical results are presented in Sec.5. The concluding remarks are given in the final section.

\section{A string model of glueballs}

According to our flux tube model \cite{IT99}, mesons (baryons) are composed of a quark and an anti-quark (diquark) connected by a color flux tube: mesons:$(q\to \bar{q})$ and baryons: $(q\to qq)$. Each arrow denotes a color flux tube. The color quantum number of $\bar{q}$ is the same as that of $qq$ so that both flux tubes are identical. This fact means the universality of the slope of the Regge trajectories of mesons and baryons. Of course, since the mass of $\bar{q}$ is different from that of $qq$, the lengths of the flux tubes are different.

Next let us consider a closed flux tube of glueballs, which are supposed to be made up of two gluons. First, we notice that the color quantum number of a gluon is the same as that of a quark pair $(q\bar{q})$ belonging to an octet representation of SU(3). Generally a $q\bar{q}$ pair is decomposed into two representations as follows:
\begin{equation}
q\bar{q}=(q\bar{q})_{1}\oplus (q\bar{q})_{8}
\end{equation}
The first term in the right-hand side denotes a singlet representation which corresponds to a meson state. On the other hand, the second term which is an octet has the same color quantum number as a gluon. Noting that a quark (anti-quark) is a source (sink) of color flux, the gluon has two flux tubes; one of them flows out and the other in. Therefore if two-gluon state is a color singlet, two flux tubes of one gluon have to connect with other two flux tubes of the other gluon to form a circle (ring) \cite{IP82}. This closed string becomes nothing but a glueball. From this formation of the color flux, it is apparent that the flux tube of the glueball is the same as those of mesons and baryons.

Now we will construct a mathematical expression for this glueball. Let us start with the MIT Lagrangian applied to a torus (bag) \cite{CJJTW74},
\begin{equation}
L=\frac{1}{2}\int ({\bf H}^{2}-{\bf E}^{2})d^{3}x -\int Bd^{3}x,
\end{equation}
where ${\bf E}$ and ${\bf H}$ are color electric and magnetic fields respectively and the $B$ in the second term is the bag constant. We consider the breathing motion of the radius of a flux ring and the rotational one arround a diameter of the ring; the closed string with varying radius $r(t)$ rotates around $x$ axis with angular velocity $\omega=\dot{\theta}(t)$ as shown in Fig.1. 
\begin{figure}
\includegraphics[width=\linewidth]{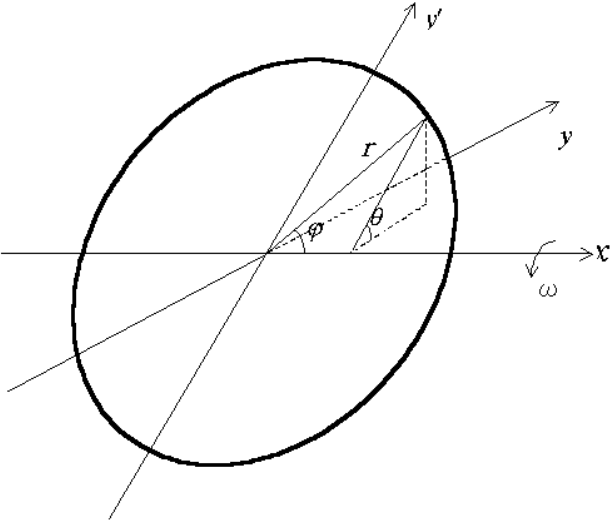}
\caption{\label{fig:epsart}The closed string with radius $r$ rotating around $x$-axis with angular velocity $\omega$.}
\end{figure}
The volume integral in the right-hand side of Eq.(2) can be rewritten in the coordinates system which rotates together with the ring. If the color flux in the rotating coordinate is denoted by ${\bf E}_{0}$, we get ${\bf H}^{2}-{\bf E}^{2}=-{\bf E}_{0}^{2}$. Then the Lagrangian is approximated by the line integral,
\begin{equation}
L(r,\dot{r},\omega)=-\frac{1}{2}\oint {\bf E}_{0}^{2}\Delta S_{0}\sqrt{1-v^2}dl-\oint B\Delta S_{0}\sqrt{1-v^2}dl
\end{equation}
where the volume element of the ring with the infinitesmal length $dl$ is approximated by $\Delta S_{0}\sqrt{1-v^2}dl$. The $\Delta S_{0}$ means cross section of the ring in the rotational coordinates system. The velocity ${\bf v}$ of the line element is perpendicular to the direction of the line element so that the factor $\sqrt{1-v^2}$ represents the Lorenz contraction. If the generalized coordinates of the line element is written as those in the Fig.1, the squared velocity is given by $v^2=\dot{r}^2+r^{2}\omega^{2}\sin^{2}\varphi$. Consequently we obtain the following Lagrangian of the closed ring,
\begin{equation}
L(r,\dot{r},\omega)=-\int_{0}^{2\pi} ar\sqrt{1-v^2}d\varphi,
\end{equation}
where the constant $a$ means string tension given by $a\equiv ({\bf E}_{0}^{2}/2+B)\Delta S_0$. Moreover if we think of the pressure balance on the flux tube, we get ${\bf E}_{0}^{2}/2=B$ \cite{JT75}. The final expression of our Lagrangian is simple. This result is realized by the following: if the closed string is divided into many particles (line elements) $\Delta m_{i}$, the total Lagrangian of this system becomes $L=-\sum_{i}\Delta m_{i}\sqrt{1-v_{i}^2}$ and this is our Lagrangian obtained above.

\section{Mass spectra of glueballs}

In this section, we quantize the Lagrangian in order to calculate the energy eigenvalues of the glueballs. The generalized coordinates of our system are the radius $r$ of the flux ring and the rotational angle $\theta$ of the ring around the $x$-axis. The conjugate momenta of these coordinates are defined by
\begin{eqnarray}
p_r&\equiv& \frac{\partial L}{\partial \dot{r}}=\int \frac{ar\dot{r}}{\sqrt{1-v^2}}d\varphi \equiv H\dot{r} \\
p_\theta&\equiv& \frac{\partial L}{\partial \omega}=\int \frac{ar^{3}\omega\sin^{2}\varphi}{\sqrt{1-v^2}} d\varphi \equiv I\omega
\end{eqnarray}
where $p_{r}$ and $p_{\theta}$ represent the radial momentum and the angular momentum of the system. It will be also shown that the quantities $H$ and $I$ introduced in the right hand side correspond to the energy and moment of inertia of the closed string respectively. The Hamiltonian is given by
\begin{equation}
H\equiv \dot{r}p_r+\omega p_{\theta}-L=\int \frac{ar}{\sqrt{1-v^2}}d\varphi 
\end{equation}
This is the energy of our system (the ring) and already appeared in Eq.(5).

Now let us go over to quantum theory. The canonical quantization leads to replacement of the momenta by the following operators:
\begin{eqnarray}
p_r&\to &\hat{p}=-i\hbar\frac{1}{r}\frac{\partial}{\partial r}r \\
p_{\theta}&\to &\hat{l}=-i\hbar\frac{\partial}{\partial \theta}
\end{eqnarray}
where the form of $p_{r}$ should be regarded as the three-dimensional radial coordinate. From this expression, the normalization of the wave function is
\begin{equation}
\int_{0}^{\infty}dr\int_{0}^{2\pi}d\varphi r^{2}|\psi(r,\theta)|^2=1
\end{equation}
Thus we obtain a Schr\"odinger equation with an energy eigenvalue $E$,
\begin{equation}
\hat{H}(r,\hat{p},\hat{l})\psi(r,\theta)=E\psi(r,\theta)
\end{equation}
Noting that $[\hat{H},\hat{l}]=0$, the wave function can be written as
\begin{equation}
\psi(r,\theta)=\frac{1}{\sqrt{2\pi}}\varphi(r)e^{il\theta/\hbar}
\end{equation}
Here $l$ is an eigenvalue of the angular momentum, which is an even integer ($l=0,2,4,\cdots$) because of the boundary condition: $\psi(r,\theta=0)=\psi(r,\theta=2\pi)$. Then the Schr\"odinger equation is reduced to the equation for the new function $\varphi(r)$ of $r$,
\begin{equation}
\hat{H}(r,\hat{p},l)\varphi(r)=E\varphi(r)
\end{equation}
To solve this equation, let us divide two cases: $l=0$ and $l\neq 0$.

\subsection{$l$=0}
In this case, we get $\omega=0$ from Eq.(6). The Hamiltonian becomes $\sqrt{\hat{p}^2+(2\pi ar)^2}$ so that the Schr\"odinger equation is reduced to
\begin{equation}
\{ \hat{p}^2+(2\pi ar)^2 \}\varphi(r)=E^2 \varphi(r)
\end{equation}
It should be noted that this equation is nothing but Schr\"odinger equation of the three-dimensional harmonic oscillator. Introducing a new function $\chi(r)$ by $\varphi(r)\equiv \chi(r)/r$, we have the Schr\"odinger equation for the one-dimensional harmonic oscillator so that we obtain the familiar eigenvalues for $E^2$,
\begin{equation}
E^2=2\pi a\hbar(2n+1)
\end{equation}
where the quantum number $n$ is an odd integer ($n=1,3,5,\cdots$) from the boundary condition: $\chi(0)=0$. This result is identical to that obtained by Koma, Suganuma and Toki \cite{KST99}. 

\subsection{$l\neq 0$}
Next we consider the case where the angular momentum does not vanish. Since the eigenvalue equation cannot be solved analytically, we make use of the WKB approximation \cite{MK69}. We put the wave function $\varphi(r)$ as 
\begin{equation}
\varphi(r)=A(r)e^{iS(r)/\hbar}
\end{equation}
where an amplitude $A(r)$ and the phase $S(r)$ are unknown real functions. Substituting it into Eq.(13), we get
\begin{equation}
\hat{H}(r,e^{-iS/\hbar}\hat{p}e^{iS/\hbar},l)A(r)=EA(r)
\end{equation}
Here we expand the left-hand side of this equation with respect to $\hbar$. The lowest order equation leads to
\begin{equation}
H(r,p,l,)=E,
\end{equation}
where $p$ stands for $\frac{dS}{dr}$. From this equation, we get the phase function $S(r)$ and the energy eigenvalue $E$; This procedure will be done in Sec.5. The next order of $\hbar$ gives 
\begin{equation}
\frac{1}{2}(\frac{\partial H}{\partial p}\hat{p}+\hat{p}\frac{\partial H}{\partial p})A(r)=0
\end{equation}
where the operators on the left-hand side have been symmetrized because it should be Hermitian. With the simple calculation, this equation can be transformed into 
\begin{equation}
\frac{d}{dr}(\frac{\partial H}{\partial p}r^{2}A)=0
\end{equation}
Therefore the amplitude function $A(r)$ is expressed by
\begin{equation}
A(r)^2=\frac{C}{\frac{\partial H}{\partial p}r^2}
\end{equation}
where the $C$ is a constant which should be determined by the normalization condition Eq.(10), that is,
\begin{equation}
C^{-1}=\int_{0}^{\infty}(\frac{\partial H}{\partial p})^{-1}dr
\end{equation}
Thus we have obtained the wave function of our system. The amplitude will be used to determine decay widths of glueballs, which is done in the next section.

\section{Decay widths of glueballs}

As mentioned in the Introduction, the existence of glueballs are predicted theoretically and has been "found" on a lattice \cite{BSHIMS93}-\cite{MP97}. It is, however, not identified experimentally. This may suggest that even if they are produced, they decay rapidly. Therefore it is very interesting to investigate the decay widths of our flux ring. To this end, we must introduce new interaction in our present model.

We assume that this new mechanism is derived by the pair creation of $q\bar{q}$: string $\to q\bar{q}\to$ many hadrons. This mechanism that is called the Schwinger one \cite{S51} was used to calculate the decay widths of mesons and baryons in Ref.(3) by us. The probability of the pair production in a unit space-time volume in the flux tube is given by
\begin{equation}
w=\frac{a^2}{4\pi^3}\sum_q \sum_{n=1}^\infty \frac{1}{n^2}\exp(-\frac{n\pi m_q^2}{a}),
\end{equation}
where $a$ is the string tention and $\sum_q$ indicates a summation over all quark flavors with mass $m_q$ (q=u,d,s) \cite{G79}-\cite{GM83}. The probability of the decay (pair production) in the time interval $dt$ is given by
\begin{equation}
dW=\int \mid \varphi(r) \mid^2 w\Omega(r)r^2drdt,
\end{equation}
where $\Omega(r)$ denotes the volume of the ring when its radius is $r$. It is given by
\begin{equation}
\Omega(r)=\Delta S_{0}\int_{0}^{2\pi}\sqrt{1-v^2}rd\varphi
\end{equation}
Substituting this equation into Eq.(24), we obtain the decay width,
\begin{equation}
\Gamma\equiv \frac{dW}{dt}=\int_{0}^{\infty}4br^{3}\sqrt{1-(\frac{p}{E})^2}K_{2}(k)A^{2}dr
\end{equation}
where $K_{2}(k)$ is the complete elliptic integral of the second kind, which is defined by
\begin{equation}
K_{2}(k)\equiv \int_{0}^{\pi/2}\sqrt{1-k^{2}\sin^{2}\varphi}d\varphi
\end{equation}
The new quantity $k$ ($0\le k<1$) is defined by
\begin{equation}
k^2\equiv \frac{r^{2}\omega^{2}}{1-(\frac{p}{E})^2}
\end{equation}
The constant $b=w\Delta S_0$ defined in the right-hand side of Eq.(26) means probability of the string breaking per unit length per unit time.

The decay width obtained above can be related to the excited energy in the case of zero angular momenta ($l=0$). First we notice $K_{2}(k=0)=\pi/2$ due to $\omega=0$. Then Eq.(26) is transformed into
\begin{equation}
\Gamma=\frac{4\pi^{2}ab}{E}\int_{0}^{\infty}A^{2}r^{4}dr=\frac{4\pi^{2}ab}{E}\langle r^{2}\rangle
\end{equation}
The $\langle r^2\rangle$ denotes the expectation value for the excited state. Therefore the decay width is simply expressed by
\begin{equation}
\Gamma=\frac{b}{2a}E,
\end{equation}
where $E$ is the energy of the corresponding excited state. The decay width is proportional to the energy in the glueball spectroscopy. This fact is natural, because both the pair production and the energy are proportional to the volume of the flux tube.

\section{Numerical results}

Since the mass spectra and the decay widths of glueballs have been given in our model, we will carry out the numerical calculations, in particular in the case of the rotational motion. First we take up the energy and the angular momentum of our system, which are constants of motion,
\begin{eqnarray}
E&=&\frac{4ar}{\sqrt{1-(\frac{p}{E})^2}}\int_{0}^{\pi/2}\frac{1}{\sqrt{1-k^{2}\sin^{2}\varphi}}d\varphi  \nonumber \\
&\equiv& \frac{4ar}{\sqrt{1-(\frac{p}{E})^2}}K_{1}(k) \\
l&=&\frac{4ar^3 \omega}{\sqrt{1-(\frac{p}{E})^2}}\int_{0}^{\pi/2}\frac{\sin^{2}\varphi}{\sqrt{1-k^{2}\sin^{2}\varphi}}d\varphi  \nonumber \\
&\equiv& \frac{4ar^3 \omega}{\sqrt{1-(\frac{p}{E})^2}}\frac{K_{1}(k)-K_{2}(k)}{k^2} 
\end{eqnarray}
where $K_{1}(k)$ is the complete elliptic integral of the first kind. If the first equation is squared, the energy is represented as
\begin{equation}
E^2=p^2+16a^{2}r^{2}K_{1}(k)^2\equiv p^2+V^2.
\end{equation}
The $V$ in the right-hand side is defined by
\begin{equation}
V(r)=4arK_{1}(k),
\end{equation}
which means a scalar potential of the closed string. Here it should be noted that the potential is not at all linear function of $r$, because the $k$ in this equation is a complicated function of $r$ as shown in Eq.(32). 

It is instructive for us to figure out the potential $V(r)$ as a function of $r$. From the other equation (32) on the angular momentum, the following relation can be derived,
\begin{eqnarray}
\frac{1}{r^2}&=&\frac{4ak}{l}\int_{0}^{\pi/2}\frac{\sin^{2}\varphi}{\sqrt{1-k^{2}\sin^{2}\varphi}}d\varphi \\
&=&\frac{4a}{kl}(K_{1}(k)-K_{2}(k)). 
\end{eqnarray}
This equation gives the function $k(r)$ of $r$ so that we are able to draw the graph of the potential $V(r)$.  The result is shown in Fig.2. 
\begin{figure}
\includegraphics[width=\linewidth]{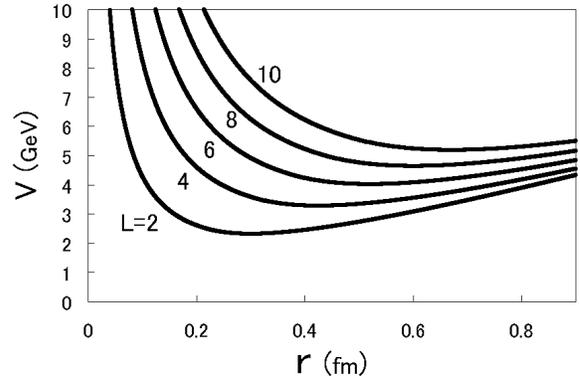}
\caption{\label{fig:epsart}The potential for the breathing mode. The $L$ in the figure denotes for the angular momentum of the closed string.}
\end{figure}
The repulsive core appears near the origin, which is due to the centrifugal potential. On the contrary, the potential becomes linear at larger distance. These behavior is natural and understood analytically as follows. Using Eqs.(34) and (36), the potential is rewritten as
\begin{equation}
V(r)=\frac{kl}{r}+4aK_{2}(k)r
\end{equation}
When $r\to 0$, we note that $k\to 1$ from Eq.(35). Then the potential is reduced to the centrifugal one: $E\to \sqrt{p^2+(l/r)^2}\approx p+l^2/(2pr^2)$. On the other hand, it approsches the linear potential at larger distance, which is the confining potential: $V(r)\to 2\pi ar$.

The excited energy can be calculated using Eq.(33). From the boundary condition of the wave function, we obtain the Bohr-Sommerfeld formula:
\begin{equation}
\oint p(r)dr=(2n+1)2\pi \hbar
\end{equation}
If Eq.(33) is substituted into the left-hand side of this equation, the energy $E$ is determined ($n=1,3,5,\cdots$). 

There are two parameters in our model: the string tension $a$ and the string breaking probability $b$. We take the same values as those used in the previous paper \cite{IT99}, which reproduce the string tension $a=0.15~{\rm GeV}^2$ and the decay width $\Gamma_{\rho}=151~{\rm MeV}$.

First let us discuss the mass spectra of the excited states. The mass spectrum of the vibrational mode given from the Eq.(15) is shown in Fig.3.
\begin{figure}
\includegraphics[width=\linewidth]{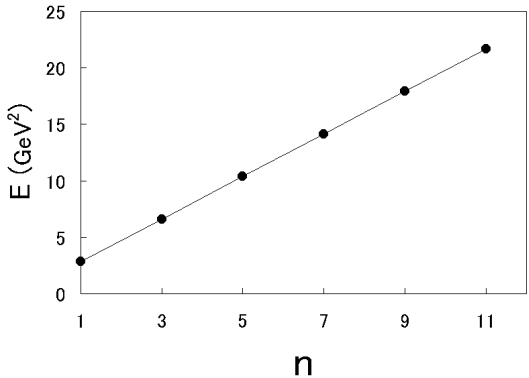}
\caption{\label{fig:epsart}The mass spectrum of the vibrational mode as a function of the vibrational quantum number.}
\end{figure}
The horizontal axis is the vibrational quantum number and the vertical one the squared mass. The quantum number must be odd integer ($n=1,3,5,\cdot$) as noted above. This straight line is an analog of the Regge trajectory of rotational motion. The vibrational modes on the straight line correspond to the l=0 states that lie on the (Regge) parent and (many) daughter trajectories with equal spacing. On the other hand, the spectrum of the rotational one calculated by Eq.(38) is shown in Fig.4. The horizontal axis is the angular momentum ($L=0,2,4,\cdots$),
\begin{figure}
\includegraphics[width=\linewidth]{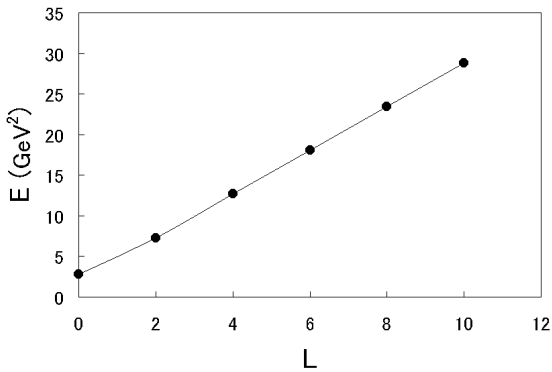}
\caption{\label{fig:epsart}The mass spectrum of the rotational mode as a function of the angular momentum.}
\end{figure}
In this case, the boundary condition at $r=0$ does not exist due to the existence of the centrifugal potential. The mass spectrum in Fig.4 is drawn in the case of $n=0$. We have obtained an almost straight line: $M^2\propto L$. The Regge slope $dL/dM^2$ is smaller than that of the ordinary Regge trajectories of mesons and baryons. This discrepancy seems to originate from the geometrical structure of the string. If the total length of the string is fixed, the moment of inertia of the ring is smaller than that of the stick-like string (meson and baryons). On the other hand, the energy of the string would be proportional to the total length. Therefore the slope of the Regge trajectory of the glueball is smaller than that of the ordinary hadrons. This problem will be discussed later.

The decay widths of the glueball are shown by the closed circles in Fig.5 and Fig.6.
\begin{figure}
\includegraphics[width=\linewidth]{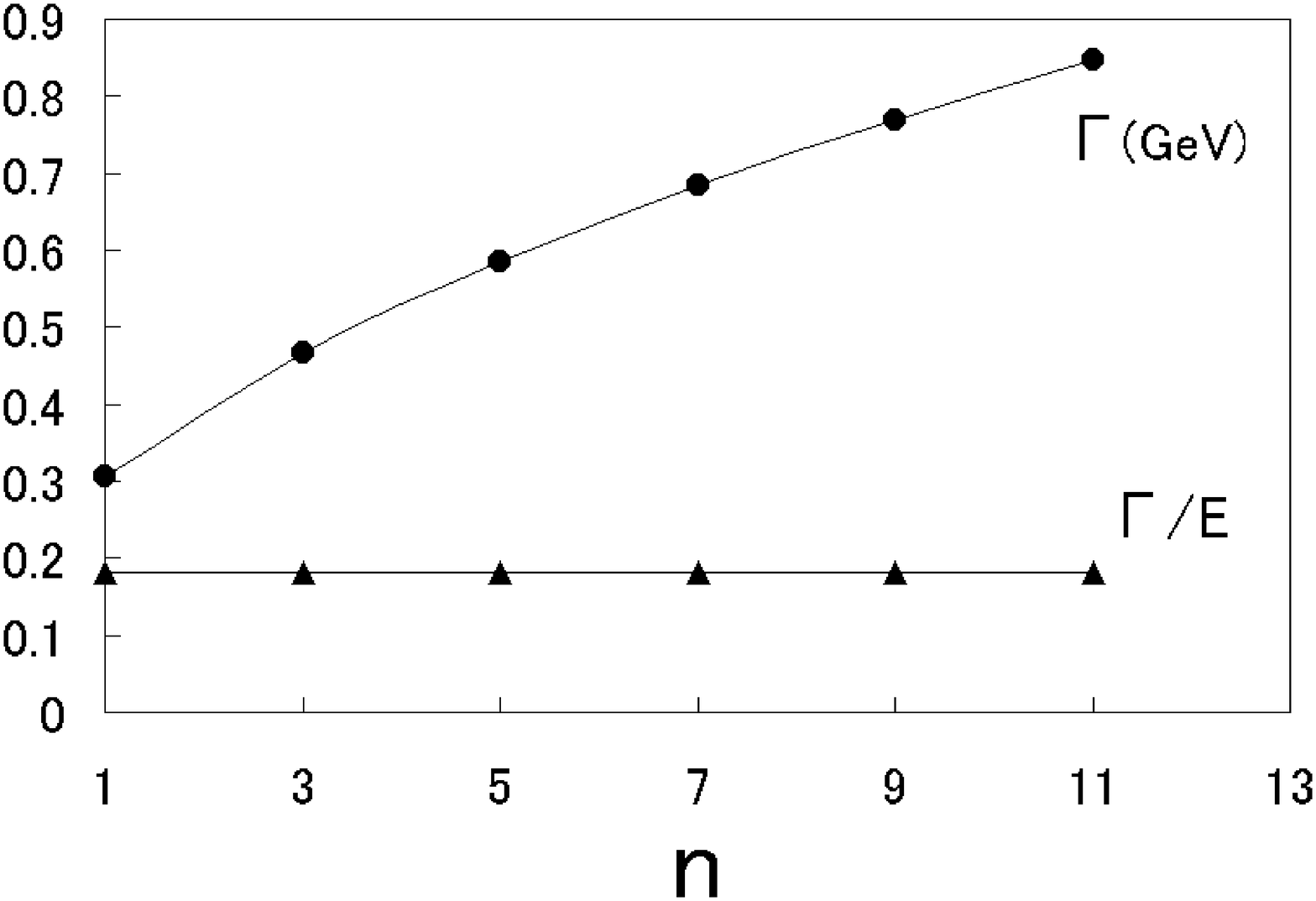}
\caption{\label{fig:epsart}The decay widths of the vibrational mode as a function of the vibrational quantum number. The triangles in the figure denote the ratio of the decay width to the mass.}
\end{figure}
\begin{figure}
\includegraphics[width=\linewidth]{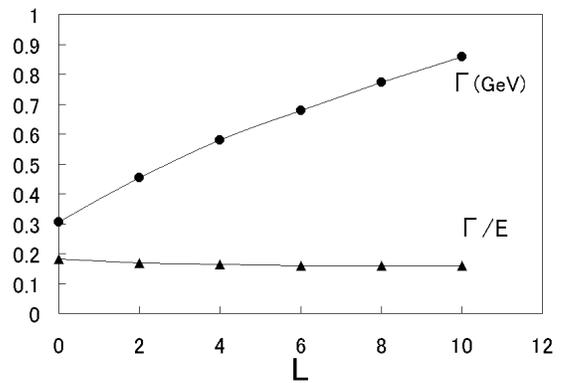}
\caption{\label{fig:epsart}The decay widths of the rotational mode as a function of the angular momentum. The triangles in the figure denote the ratio of the decay width to the mass.}
\end{figure}
They are rather larger than those of the ordinary hadrons so that the glueballs generally decay easily. This fact explains that no glueball have been found experimentally. On the other hand, recent lattice QCD simulations suggest that the decay width of the scalar glueball is narrow \cite{BP97,CHW98,MP97}, which is caused on the OZI rule. If the probability of the decay to one $q\bar{q}$ pair state is denoted by $w_1$, we should use the $w-w_1$ instead of $w$ in our above calculations due to the OZI rule. Therefore the resulting decay rate is lowered. The triangles in the same figures represent the ratio of the decay width to the mass. In the case of the vibrational mode (Fig.5), the decay widths are proportional to their mass. This is an exact result as shown in Eq.(30). The constancy of the ratio seems to hold also for the rotational mode as seen in the Fig.6. 

The relation between the decay width and the mass is understood as follows. The mass of hadrons is proportional to the length of the string. On the other hand, the breaking probability is also proportional to the length of it because the cross section of the tube is the same for all hadrons (universality). Therefore we obtain the linearity: $\Gamma\propto E$.

Finally we add some comments on the experimental data \cite{B96}. There are some candidates: $f_{0}(1500)$ and $f_{0}(1710)$ for the scalar glueball; $f_{2}(2300)$ and $f_{2}(2340)$ for the tensor glueball. Unfortunately the abundance of $q-\bar{q}$ meson states in the 1-3 GeV region and the possibility of quarkonium-glueball mixing states still make it difficult to identify the glueball states. Instead we refer results by the lattice QCD simulation \cite{MP99}, which is shown in Table I. 
\begin{table}
\caption{Mass spectrum of the blueballs. They are calculated by the lattice QCD and our model.
\label{table1}}
\begin{center}
\tabcolsep=5mm
\begin{tabular}{c|cc}\hline\hline
 & lattice QCD (MeV) & Our model (MeV) \\ \hline
$0_{1}$ & 1.7 & 1.68 \\
$0_{2}$ & 2.5$\sim 2.8$ & 2.57 \\
$2_{1}$ & 2.4 & 2.69 \\
\hline\hline
\end{tabular}
\end{center}
\end{table}
The first and second $J=0$ states are expected to be the breathing modes of the glueball. These values are consistent with those by our model. The first $J=2$ state is the glueball state by the rotational motion and the energy is slightly smaller than that by our model. This is caused on the assumption that the shape of the ring is always circular. The centrifugal force would deform the shape so that the ring becomes elliptic. Taking this effect into account, the moment of inertia of the ring becomes large and the excitation energy is lowered.

\section{Concluding remarks}

We have developped the flux tube model of glueballs in the context of that of mesons and baryons. The glueball is described by the closed string whose properties are the same as those of the stick-like string that constitutes a $q-\bar{q}$ meson or a $q-(qq)$ baryon. Our model is characterized only by two parameters, the string tension $a$ and the string breaking parameter $b$. If the two parameters are taken so as to reproduce the mass and the decay width of the rho meson, our model has no free parameter. Then we have investigated the mass spectra and the decay widths of glueballs with the use of the flux tube ring model. The numerical results are consistent with those of the lattice QCD. The calculated decay widths are sevral hundreds MeV. These values are rather large so that it may be difficult to identify the glueball experimentally.

\begin{acknowledgments}
We would like to thank Nuclear Theory Group at Kochi University for helpful discussions. In particular we would like to thank to Mr. Yuma Harada for his help of numerical calculations in the early stage of this work.
\end{acknowledgments}

\end{document}